\documentclass[preprint,12pt]{elsarticle}
\usepackage{amsfonts}
\usepackage[utf8]{inputenc}
\usepackage{lineno}
\usepackage{amsmath}
\usepackage{multirow}
\usepackage[abs]{overpic}
\usepackage{lscape}

\begin{document}

\begin{frontmatter}

\title{Perennial life-histories and demographic advantages may play contradictory roles in the evolution of plant mating systems}

 \author[EEP,agro]{D.~Abu Awad\corref{cor1}}
 \ead{diala.abu-awad@supagro.inra.fr}

 \author[EEP]{S.~Billiard}
 \author[PP]{V.C.~Tran}
 \cortext[cor1]{Corresponding author}

 \address[EEP]{Unité Evolution, Ecologie et Paléontologie, UFR de Biologie, UMR 8198, CNRS / Université de Lille – Sciences et Technologies, 59655 Villeneuve d’Ascq Cedex, France}
 \address[agro]{INRA, UMR AGAP, 2 place Pierre Viala F-34060 Montpellier Cedex 1, France}
 \address[PP]{Laboratoire Paul Painlev\'e, UFR de Math\'ematiques, UMR 8524, CNRS / Université de Lille – Sciences et Technologies, 59655 Villeneuve d’Ascq Cedex, France}


\begin{abstract}
When predicting the fate and consequences of recurring deleterious mutations in self-fertilising populations most models developed make the assumption that populations have discrete non-overlapping generations. This makes them biologically irrelevant when considering perennial species with over-lapping generations and where mating occurs independently of the age group. Previous models studying the effect of perennial life-histories on the genetic properties of populations in the presence of self-fertilisation have done so considering age-dependent selection and have found that, contrary to empirical observations, perennial populations should exhibit lower levels of inbreeding depression. Here we propose a simple deterministic model in continuous time with selection at different fitness traits and feedback between population fitness and size. We find that a perennial life-history can result in high levels of inbreeding depression in spite of inbreeding, due to higher frequencies of heterozygous individuals at the adult stage. We also propose that there may be demographic advantages for self-fertilisation that are independent of reproductive success. 
\end{abstract}
\end{frontmatter}

\section{INTRODUCTION}

\par The prevalence of outcrossing in phylogenies where self-fertilisation has evolved independently several times (and with no detected returns to outcrossing \cite{IgicBusch13}) is a long-running question in evolutionary biology. Though it has been suggested that self-fertilisation is an evolutionary dead-end \cite{Takebayashi2001,Wright13}, with self-fertilising lineages suffering from higher extinction rates \cite{Goldberg10}, the short-sightedness of natural selection makes it seem unlikely that only long-term disadvantages are responsible for the maintenance of outcrossing. Short-term disadvantages or barriers to the spreading of self-fertilisation that can explain empirical observations must therefore exist. 
\par The most evident disadvantage of self-fertilisation is that of the cost of inbreeding \cite{ChCh87} leading to the expression of deleterious recessives and hence decreasing the fitness of selfed offspring. It is generally accepted that for a level of inbreeding depression lower than $0.5$, Fisher's automatic advantage \cite{Fisher41} will favour the spreading of an allele promoting self-fertilisation. As self-fertilisation increases, this should lead to a purge of deleterious alleles \cite{Glemin03}, lowering the observed level of inbreeding depression. Empirical results on the other hand have found that inbreeding depression may remain relatively high in self-fertilising species \cite{Winn11} and an analysis comparing several works on the influence of self-fertilisation on inbreeding depression suggests that perennials seem to be less efficient at purging their genetic load compared to annual species \cite{ByersWaller99}. Indeed, contrary to predictions from discrete-time models with non-overlapping generations \cite{ChCh87, Glemin03}, in the presence of self-fertilisation perenniality seems to be associated with higher than expected heterozygocity at loci under selection \cite{Spigler2009}. These observations may be key in understanding the correlation between the rate and prevalence of self-fertilisation and plant life-histories, with many annuals being self-fertilising whereas perennials tend to remain outcrossing (see \cite{Morgan97}). It has been suggested that this is because perennials maintain high life-time reproductive assurance, though a model by Morgan et al. \cite{Morgan97} supports that the maintenance of outcrossing in perennials is more likely due to the avoidance of adult inbreeding depression since self-fertilisation offers similar advantages in reproductive success for both annuals and perennials.
\par Indeed most conventional population genetics models studying the consequences of the evolution of self-fertilisation have done so assuming discrete-time and non-overlapping generations \cite{LandeSch85, Ch90,PorcherLande05, GlemRon2013}. Such models therefore neglect the potential effect of life-history traits on the maintenance of inbreeding depression. Previous works on the relationship between life-histories and self-fertilisation have found that perenniality may actually facilitate the evolution to high selfing rates due to lowered levels of inbreeding depression at later stages \cite{Morgan2001}. However,  Morgan's \cite{Morgan2001} model considered selection to occur on longevity, with each generation presenting the opportunity for less fit genotypes to be purged from the population. Though this is not an implausible definition of fitness, it does not take into account other possible components of fitness that can and do contribute to both mean fitness and inbreeding depression. 
\par The concept of fitness is essential in any work determining how natural selection influences a population's genetic and demographic state. In very general terms, the consensus is that individual fitness is it's ability not only to survive but also to reproduce in a given environment \cite{Orr2009}. In population genetics models, this remains true, though the exact definition of fitness may vary depending on how selection has been introduced (see \cite{Haldane24}, Chapter 7 of \cite{roughgarden1979} and \cite{Parsons2010} for examples of how fitness was accounted for). The very widely used discrete models with non-overlapping generations make biological assumptions that cannot be universally applied and mainly represent annual plant populations \cite{NagylakiCrow74}. In such models definitions of fitness i.e. survival or fecundity are interchangeable \cite{Haldane24,Bodmer65}. As individuals are present for only one generation this does indeed seem plausible (if an individual does not survive to reproduce or simply does not reproduce the outcome is the same), but what of populations that are not as compartmentalised as annual plants? Do different components have the same influence population equilibria as in the discrete case? 
\par When modelling perenniality, two approaches can be considered: discrete-time age-structured models (see \cite{ChAgeStructure}) or continuous-time models  (\cite{Fisher30}, Chapter 5.3 of \cite{CrowKim}, and \cite{NagylakiCrow74}). In the latter case, age-structure is of little importance if selection is considered to be age-independent as genic and genotypic frequencies in a population evolve towards time-persistent forms independently of the initial conditions considered \cite{Vlad88}. The use of continuous-time models without age-structure provides a simpler alternative  to modelling overlapping generations, there are however two main consequences: the effect of age on reproductive capacity and survival is ignored and, more generally, individuals have an indefinite life-span. Since, however, how age influences fecundity and survival is not clear-cut (see \cite{baudisch2008}), considering age-independent selection is not wholly unrealistic. 
\par Most continuous-time models stem from Fisher's Fundamental Theorem of Natural Selection \cite{Fisher30} wherein he introduced the notion of Malthusian fitness, defined as the growth rate. However, simultaneously introducing mutation, selection and non-random mating in such models can prove to be challenging and simpler models in continuous time seem to be lacking. For there to be demographic equilibrium, all genotypes must have the same Malthusian fitness, making it an inappropriate indicator of differences between genotypes in this scenario \cite{Charlesworth70}. Other definitions of fitness are therefore to be preferred when examining population equilibrium, all the more so if they facilitate the comparison to definitions of fitness in discrete-time models.
\par Another aspect that is often ignored in conventional population genetics models is that of the demographic consequences of the genetic load. Works that have addressed the potential genotypic effects of deleterious mutations on population size have done so by considering density-dependent selection, usually with a trade-off between r- and K-selection (for example \cite{Charlesworth71, Clarke72,roughgarden1979}, but in \cite{Charlesworth71} see section on density-independent selection). And in cases where mutations are unconditionally deleterious, the ecological and genetic aspects of the models were dissociated (\cite{Clarke73, AgraWhit2012}, but see the extreme case of the mutational meltdown \cite{Lande94, Lynch95}). By considering that the ecological and genetic properties vary independently, any potential feed-back between the two may be missed.That different components of fitness have different effects on population size has been suggested by different models \cite{Clarke73, AgraWhit2012}, but could the mating system further influence the consequences of selection on population demography?
\par Here we introduce a simple model in continuous time where both the demographic and genetic equilibria are emerging properties and not pre-defined parameters. We study how the rate of self-fertilisation, in interaction with different components of fitness, influences population size and the genetic properties of populations at mutation-selection balance. We compare our results to expectations from conventional population genetics models in discrete time so as to evaluate whether perenniality may play a role in the maintenance of outcrossing.

\section{Model}
\par We consider the evolution of a population with a varying population size and a single bi-allelic locus, where $A$ is the wild type and $a$ the mutant allele. The population is made up of sexually reproducing hermaphrodite individuals, who self-fertilise at a fixed rate $\alpha$ (with $\alpha = 0$ being panmictic and $\alpha = 1$ strictly self-fertilising). The environment is stable, and the population is isolated and spatially unstructured. Three genotypes can be found in the population, $aa$, $Aa$ and $AA$, which, from here onwards, are denoted $X$, $Y$, and $Z$ respectively. At a given time $t$, the population is made up of three kinds of individuals, $X_t$, $Y_t$ and $Z_t$ representing the number of individuals carrying the respective genotype. We denote the population size $N_t = X_t + Y_t + Z_t$. In a large population setting, these quantities can be considered as continuous, and the evolution of the number of individuals of each genotype is described in continuous time using ordinary differential equations. Three processes affect the change in the number of individuals of each genotype, births (occurring with rate $R_t^V$, where $V$ can be either $X$, $Y$ or $Z$), deaths (at a rate $M^V_t$) and mutation. Selection and density dependence are introduced in these processes. We consider that the mutation from $A$ to $a$ is unidirectional and occurs with a probability $\mu$ at the gamete stage. 
\par We first introduce the demographic properties of the model without considering mutation and selection and show that this model respects the genotypic frequencies predicted by the Hardy-Weinberg model for neutral alleles and that in the case of self-fertilisation, the deviation from the Hardy-Weinberg equilibrium is as expected from conventional population genetics models (\emph{i.e.} as a function of the inbreeding coefficient $F_i = \frac{\alpha}{2-\alpha}$). Mutation and selection are then introduced, with selection influencing different components of fitness (during different moments of the life cycle, but independent of age) and we define demographic and genetic variables to quantify the effect of the recurrent introduction of deleterious mutations on populations. 
\subsection*{General Model}
The equation describing the change in the number of individuals for each genotype is given by

\begin{equation}
\label{all_simple}
\frac{dV_t}{dt} = R^V_t - M^V_t ,
\end{equation}

$R^V_t$ and $M^V_t$ being the rates of birth and mortality respectively of individuals $V$, and $V$ representing $X$, $Y$ or $Z$. 
\par For each of the genotypes these birth rates $R^V_t$ are given by 

\begin{equation}
\label{modG}
\begin{split}
R^X_t &= b \left(\alpha \left(X_t + \frac{1}{4} Y_t  \right) + \frac{(1-\alpha)}{N_t} \left(X_t^2+X_t Y_t +\frac{1}{4} Y_t^2  \right)\right)\\
R^Y_t &= b \left(\alpha\left(\frac{1}{2} Y_t \right)+ \frac{(1-\alpha)}{N_t} \left( X_t Y_t +2 X_t Z_t +\frac{1}{2} Y_t^2 +Y_t Z_t  \right)\right)\\
R^Z_t &= b \left(\alpha\left( \frac{1}{4} Y_t  + Z_t \right)+\frac{(1-\alpha)}{N_t} \left(\frac{1}{4} Y_t^2 +Y_t Z_t + Z_t^2  \right)\right)
\end{split}
\end{equation}

The assumptions made in formulating the expressions for $R^V_t$ are as follows. Individuals within the population are hermaphroditic and can contribute both via the male and the female functions. Female gametes are limited, and depend on the number of individuals present in the population, whereas male gametes are produced in very large quantities (\emph{i.e.} there is no pollen limitation) and are subjected to competition.
In sum, the birth/recruitment rate depends on an intrinsic reproductive rate $b$ (which, by default, holds the same value for all genotypes) and on the reproductive events that lead to the production of new individuals with genotype $V$. 

\par The death rate $M^V_t$ depends on an intrinsic death rate $d$ and is density dependent (we consider a carrying capacity $K$). The equation for $M^V_t$ is given by

\begin{equation}
\label{mort}
M^V_t = d \frac{N_t}{K}V_t .
\end{equation}

The choice of density dependence on mortality is arbitrary as a symmetrical form of density dependence on birth, written in the form $b\frac{K}{N_t}$ yields similar results as those presented below. When solving $\frac{dN}{dt} = \frac{dX}{dt} = \frac{dY}{dt} = \frac{dZ}{dt} = 0$ we find the optimal population size is given by (see Supplementary Material S4 for the proof)

\begin{equation}
\label{Neqn}
N_{eq} = \frac{b K}{d} .
\end{equation} 

\par By solving the equations we find that the frequencies of $X$, $Y$ and $Z$ concord with the expectations of the generalised Hardy-Weinberg law, with genotypic frequencies depending on the inbreeding coefficient $F_i =\frac{\alpha}{2-\alpha}$ ( see p.120 of \cite{gillesp} and Supplementary Material S3). When the rate of self-fertilisation $\alpha = 0$, then $F_i = 0$ and the frequencies of $X$, $Y$ and $Z$ are at Hardy-Weinberg equilibrium.

\subsection*{Introducing Mutation}
\par Mutations occur at a rate $\mu$ during gamete formation and are considered to be uni-directional from $A$ to $a$. The proportions of $a$ gametes produced per genotype are therefore $1$, $\frac{1 + \mu}{2}$ and $\mu$ for $X$, $Y$ and $Z$ individuals respectively (the proportions of $A$ alleles being $0$, $\frac{1 - \mu}{2}$ and $1-\mu$). Mutational events are integrated into the birth rate $R^V_t$, with, for example :

\begin{equation}
\label{mutX}
\begin{split}
R^X_t &= b \bigg(\alpha \left(X_t + \frac{1}{4} Y_t (1 + \mu)^2 + Z_t \mu^2 \right) \\ &+ \frac{(1-\alpha)}{N_t} \bigg(X_t^2+X_t Y_t (1+\mu )+2 X_t Z_t \mu+\frac{1}{4} Y_t^2 (1+\mu )^2\\ & \hskip15em+Y_t Z_t \mu (1+\mu) +Z_t^2 \mu ^2\bigg)\bigg).
\end{split}
\end{equation}

\subsection{Timing of selection\\}
\par
\begin{figure}
\center
\includegraphics[width=0.9\textwidth, clip, trim = 2em 2em 2em 2em]{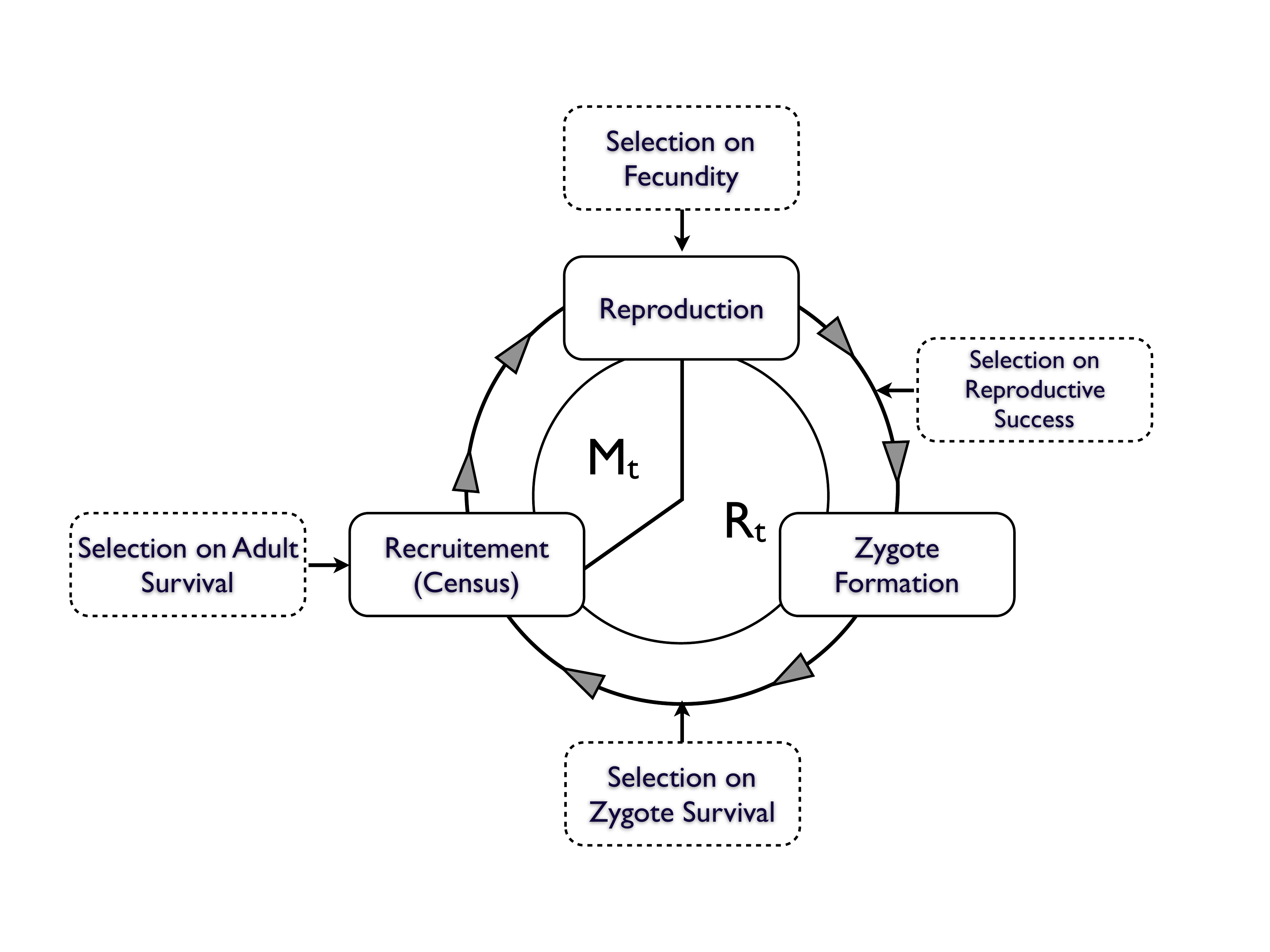}
\caption{\small{A schematic representation of the life-cycle modelled. Selection on the components of fitness are represented in boxes with dashed borders. $R_t$ and $M_t$ represent the birth and death functions respectively. Individuals, once recruited, remain a part of the censused population, contributing offspring (at rate $R_t$) until their death (at rate $M_t$).  }}
\label{cycle}
\end{figure}

 We consider selection that is not age-dependent and influences the population's net reproductive rate ($R_t - M_t$). As the genetic properties at equilibrium are emerging properties of the model, we examine the potential effects of the fitness component in the presence of non-random mating on the allelic and genotypic frequencies at mutation-selection balance. Selection can either affect reproduction (relative reproductive success or fecundity) or survival (at either the zygote or the adult stage).  The deleterious allele $a$ has a coefficient of selection $s$ and dominance $h$, giving a relative fitness at a given trait of $(1 - s)$, $(1 - h s)$ and 1 for genotypes $X$, $Y$ and $Z$ respectively. Below we detail the hypotheses made when defining the different fitness components. The full equations for the change in the number of individuals of each genotype for these models can be found in Supplementary Material S2. A summary of the life-cycle and the different forms of selection can be seen in Figure \ref{cycle}.

\subsubsection{Selection on reproduction: }
In order to model the effect of the deleterious allele $a$ on the reproductive output of individuals, we introduce a new term $\widetilde{V}_t$ instead of $V_t$ in the $R_t^V$ function. This term represents the contribution of $V$ individuals to the genetic pool (or their reproductive output), proportional to their fitness (i.e. $\widetilde{X}_t = (1-s) X_t$). Carrying $a$ can influence the reproductive output by either reducing the reproductive success of individuals or their fecundity.

\paragraph{\it Reproductive Success (model RS): }
When reproductive success is reduced, all individuals produce the same quantity of gametes, however the success of a mating event depends on the genotypes' fitness (\emph{i.e.} for $X$ individuals only $(1-s)$ matings lead to fertilisation). The proportion of female and male gametes an individual $V$ effectively contributes are therefore $b \widetilde{V}_t$ and $\frac{\widetilde{V}_t}{N_t}$. $R^X_t$ for this model of selection is given by

\begin{equation}
\label{mod1}
\begin{split}
R^X_t =& b\bigg(\alpha\left(\widetilde{X}_t + \frac{1}{4} \widetilde{Y}_t (1+\mu)^2 + \widetilde{Z}_t \mu^2 \right)\\& + \frac{(1-\alpha)}{N_t} \bigg(\widetilde{X}_t^2+\widetilde{X}_t \widetilde{Y}_t (1+\mu)+2 \widetilde{X}_t\widetilde{Z}_t \mu+\frac{1}{4} \widetilde{Y}_t^2 (1+\mu )^2 \\& \hskip10em+ \widetilde{Y}_t \widetilde{Z}_t \mu (1+\mu)+\widetilde{Z}_t^2 \mu ^2 \bigg)\bigg) .
\end{split}
\end{equation}

\paragraph{\it Fecundity (model F): }
Selection reducing individual fecundity is translated by an individual $V$ contributing $b \widetilde{V}_t$ female gametes and a proportion of $\frac{\widetilde{V}_t}{\widetilde{X}_t +\widetilde{Y}_t +\widetilde{Z}_t}$ male gametes to the next generation. The equations for $R^V_t$ with selection on fecundity are therefore the same as for selection on reproductive success, with the exception of the term $\frac{(1-\alpha)}{N_t}$, which is replaced by $ \frac{(1-\alpha)}{\widetilde{X}_t +\widetilde{Y}_t +\widetilde{Z}_t}$. Note that for this form of selection the proportion of male gametes effectively contributed by $V$ depends on the total amount of male gametes produced and not on the number of individuals in the population as with selection on reproductive success. 

\subsubsection{Selection on survival: }
Selection can also occur during the life cycle, independently of reproductive output, affecting either zygote or adult survival. 
\paragraph{\it Zygote Survival (model Z): }
Zygote survival can be translated as the proportion of viable offspring that are recruited into the population. If $V$ has a probability of being recruited proportional to its fitness, this implies that the effective reproductive output $R^V_t$ depends on $V$'s relative fitness $W^V$. The function $R^V_t$ is therefore multiplied by the genotypic fitness, with, for example :

\begin{equation}
\label{mod3}
\frac{dX}{dt} = (1-s) R^X_t  - M^X_t  
\end{equation} 

One can note that the symmetrical equivalent of this model, with selection increasing mortality (done by replacing $d$ in $M^V_t$ by $\frac{d}{W^V}$), leads to the same results.

\paragraph{\it Adult Survival (model A): }
For this form of selection we make the assumption that adults are eliminated before they have reproduced and have little effect on density dependence (i.e. that the resources $K$ consumed by the these adults is negligible). As adults that are selected against neither contribute to the next generation, nor generate competition for male reproduction the expression for $R^V_t$ in this model of selection is equivalent to that for selection on fecundity (equation \ref{mod1} with $\frac{(1-\alpha)}{N_t}$, replaced by $ \frac{(1-\alpha)}{\widetilde{X}_t +\widetilde{Y}_t +\widetilde{Z}_t}$). similarly, the assumed lack of competition for resources leads to a new expression for the death rate $M^V_t$, given by

\begin{equation}
\label{mod3b}
M^V_t = d \frac{\widetilde{X}_t+\widetilde{Y}_t+\widetilde{Z}_t}{K} V_t .\\
\end{equation}
\vskip1em

\subsection{Population equilibrium}
\par The deterministic equilibrium values for each of the models of selection described above (models RS, F, A and Z) are derived by solving $\frac{dX_t}{dt} = \frac{dY_t}{dt} = \frac{dZ_t}{dt} = 0$. This allows us to obtain the number of individuals carrying each genotype at equilibrium ($X_{mut}$, $Y_{mut}$ and $Z_{mut}$), the sum of which gives us the population size at equilibrium ${N}_{mut}$. In cases where no explicit solution could be found, numerical iterations were performed to obtain the numbers of each genotype at equilibrium.
The expressions for $X_{mut}$, $Y_{mut}$ and $Z_{mut}$ are then used to derive the expressions for the mutational load $L$ and inbreeding depression $\delta$. The mutational load $L$ is defined as the decrease in mean fitness due to the presence of deleterious mutations and is given by p.61 of \cite{gillesp}:

\begin{equation}
\label{load}
L = 1 - \frac{(1-s) X_{mut} + (1 - hs) Y_{mut} + Z_{mut}}{N_{mut}} \quad .\\
\end{equation}

\par We also explore the expected level of inbreeding depression $\delta$ in populations, which is calculated using equation $3$ in \cite{RR2004}:

\begin{equation}
\label{del}
\delta= 1 - \frac{W_s}{W_o}
\end{equation}

where $W_s$ is the fitness of selfed offspring and $W_o$ of outcrossed offspring and are given by

\begin{equation}
\begin{split}
W_s &= (1-s) X_{mut} + \left(\frac{1}{4} + \frac{1-hs}{2} + \frac{1-s}{4} \right)Y_{mut} + Z_{mut} .\\
W_o &= (1-s) (X_{mut}+ \frac{Y_{mut}}{2})^2\\&+ (1 - hs) (X_{mut}+ \frac{Y_{mut}}{2})(Z_{mut}+ \frac{Y_{mut}}{2}) + (Z_{mut}+ \frac{Y_{mut}}{2})^2 .
\end{split}
\end{equation}

We then compare $L$ and $\delta$ to expectations from conventional population genetics models. In order to compare our results to these models, we replace $X_{mut}$, $Y_{mut}$ and $Z_{mut}$ with $q^2$, $2q(1-q)$ and $(1-q)^2$ respectively, where $q$ is the frequency of the deleterious mutant $a$ at mutation-selection balance. We will compare our models to the explicit expression for $q$ (for any value of $s$, $\mu$ and $\alpha$, but $h \neq 0.5$) from the model presented in Chapter 6 of \cite{CrowKim}, where

\begin{equation}
\label{qck}
q^{CK}_{\alpha,h} =\frac{\sqrt{G_{\alpha,h}}-s (h (1+\mu ) (1-F_i)+F_i)}{2 (1-F_i) (1-2 h) s}
\end{equation}

and

\begin{equation}
\label{G}
G_{\alpha,h} =s \left(4 \mu  (1-F_i) (1-2 h)+s (F_i+(1-F_i) h (1+\mu ))^2\right) .
\end{equation}

Using these expressions we can obtain expressions for the genetic load $L^{CK}_{\alpha,h}$ and the level of inbreeding depression $\delta^{CK}_{\alpha,h}$
\par In order to provide multi-locus estimations, if the explicit equations are available, then mean fitness is calculated using Haldane's \cite{Haldane27} approximation for multiplicative independent loci, where fitness $W \approx Exp[-L]$ and the single locus mutation rate $\mu$ in $L$ is replaced by the genomic mutation rate. In the case of numerical iterations, then $W = (1 - L)^{n}$, $n$ representing the number of loci and $\mu$ being maintained as the single locus mutation rate. In order to estimate the level of inbreeding depression $\delta$, we use Kirkpatrick and Jarne's \cite{KirkpatrickJarne2000} expression where inbreeding depression due to multiple loci is $$\Delta = 1- \prod_{i = 1}^{n}(1-\delta_i) .$$
\par In cases where the genotypic frequencies from our model deviated from theoretical Hardy-Weinberg proportions for a given frequency of the deleterious allele at equilibrium $q$, two measures are made in order to verify the consequences of this deviation: $1)$ Wright's Inbreeding coefficient, $F_W$, calculated using 
\begin{equation}
\label{FW}
F_W = 1 - \frac{Y_{obs}}{2q(1 - q)}\quad,
\end{equation}

where the numerator ($Y_{obs}$) and denominator are the frequency of heterozygotes obtained from our model and that expected for a given frequency of the deleterious allele $q$ in the absence of inbreeding respectively and $2)$ A modified expression of Wright's coefficient of inbreeding
\begin{equation}
F_{excess} = 1 - \frac{Y_{obs} }{2q(1 - q)(1 - F_i)}\quad,
\end{equation}
\label{Fexcess}  
where $Y_{obs}$ is as before and the denominator is the expected frequency of heterozygotes from conventional models in the presence of inbreeding.

\par For all four models of selection, there exists a solution where the population is made entirely of $X$ individuals. There is therefore a threshold value of the mutation rate $\mu$, as a function of the selection coefficient $s$ and the dominance $h$, which leads to the deterministic fixation of $a$. However, as this requires that the selection coefficient $s$ must be of order $\mu$, $a$ would be selectively neutral and drift the main force influencing its frequency, rendering this threshold biologically irrelevant. We thus do not consider this case further, but more detail can be found in Supplementary Material S1.

\section{Results}
\par By solving the equations given in the previous section, we have found explicit solutions when the population is panmictic ($\alpha = 0$). Clear explicit solutions for any rate of self-fertilisation were possible only for recessive mutations ($h = 0$). For other values of $h$ ($\neq 0.5$) with $\alpha \neq 0$, solutions could not be obtained and so the variables at equilibrium were calculated using numerical iterations. All mathematical operations were carried out using Wolfram's Mathematica $9.0$ \cite{mathematica}. 

\begin{landscape}
\begin{table}[htbp]
\center
{\renewcommand{\arraystretch}{1.8}
\begin{tabular}{|c|c|c|c|c|c|c|}
\cline{1-7}
\multirow{2}{*}{Model}&\multicolumn{2}{c|}{$\alpha = 0$ \& $h \neq 0.5$}&\multicolumn{4}{c|}{$\alpha \neq 0$ \& $h = 0$}\\ [0.2cm]\cline{2-7}
&$q$&$L$&$q$&$L$&$X_F$&$\delta$\\ [0.2cm]\cline{1-7}
CK&\multirow{4}{*}{$q^{CK}_{0,h}$}&\multirow{4}{*}{$L^{CK}_{0,h}$}&$\frac{\mu  (2- \alpha )}{s \alpha } + o(\mu )^2$&\multirow{4}{*}{$\mu$}&\multirow{4}{*}{$\frac{\mu}{s}$}&$\frac{\mu(2-\alpha )}{2 \alpha }+o(\mu )^2$\\ [0.2cm]\cline{1-1}\cline{4-4}\cline{7-7}
RS&&&\multirow{3}{*}{$\frac{\mu  (2-(1-2 s) \alpha )}{s \alpha }+o(\mu )^2$}&&&\multirow{3}{*}{$\mu\left(\frac{1+\alpha s}{\alpha }\right)+o(\mu )^2$}\\ [0.2cm]\cline{1-1}
F&&&&&&\\ [0.2cm]\cline{1-1}
A&&&&&&\\ [0.2cm]\cline{1-7}
&\multicolumn{2}{c|}{$h = 0$}&&&&\\\cline{2-3}
Z&$\frac{\sqrt{\frac{\mu }{s}}-\mu }{1-\mu }$&$\frac{\mu (1 - s)}{1 - \mu}$&$\frac{\mu  (2-(1-s) \alpha )}{s \alpha }+ o(\mu )^2$&$\frac{\mu (1 - s)}{1 - \mu}$&$\frac{\mu  (1-s)}{(1-\mu ) s}$&$\frac{\mu }{\alpha }+ o(\mu )^2$\\ [0.2cm]\cline{1-7}
\end{tabular}
}\caption{Effect of the mode of selection on genetic load $L$ and frequency $q$ of the deleterious allele for a panmictic population ($\alpha = 0$) and all values of $\mu$, $s$ and $h \neq 0.5$, and $L$, $q$, the frequency of the deleterious homozygote $X_F$ and the level of inbreeding depression $\delta$ for a population with selfing rate $\alpha \neq 0$ and $h = 0$. In the latter case $q$ and $\delta$ are obtained to the first order of $\mu$ from the explicit analytic results, making it valid only for very small values of $\mu$ and $\alpha>> 0$ (Full equations can be obtained using the expressions for the frequencies $X_F$, $Y_F$ and $Z_F$ in the appendix). $q^{CK}_{0,h}$ is given by setting $F_i = 0$ in equation \ref{qck}, and is then used to calculate $L^{CK}_{0,h}$ as shown in equation \ref{load}. Full equations of genotypic frequencies for recessive mutations can be found in Supplementary Material S3.} 
\label{Lmut}
\end{table}
\end{landscape}

\subsection{Allelic frequencies and the genetic load}
\par In the absence of self-fertilisation, selection models RS, F and A result in the same allelic frequencies and genetic load $L$ as expected from conventional population genetics models (see Table \ref{Lmut}). Model Z differs in that the frequency $q$ at equilibrium is that obtained from conventional models in discrete time after selection and before mutation have taken place. This is intuitive in that census occurs at the adult stage, hence after unfit zygotes have been eliminated. The strength of selection therefore has a greater effect on $L$, whereas for the other models, when mutations are recessive $L = \mu$. This however is of little consequence numerically.

\par
\begin{figure}
\center
\begin{overpic}[scale=0.81]{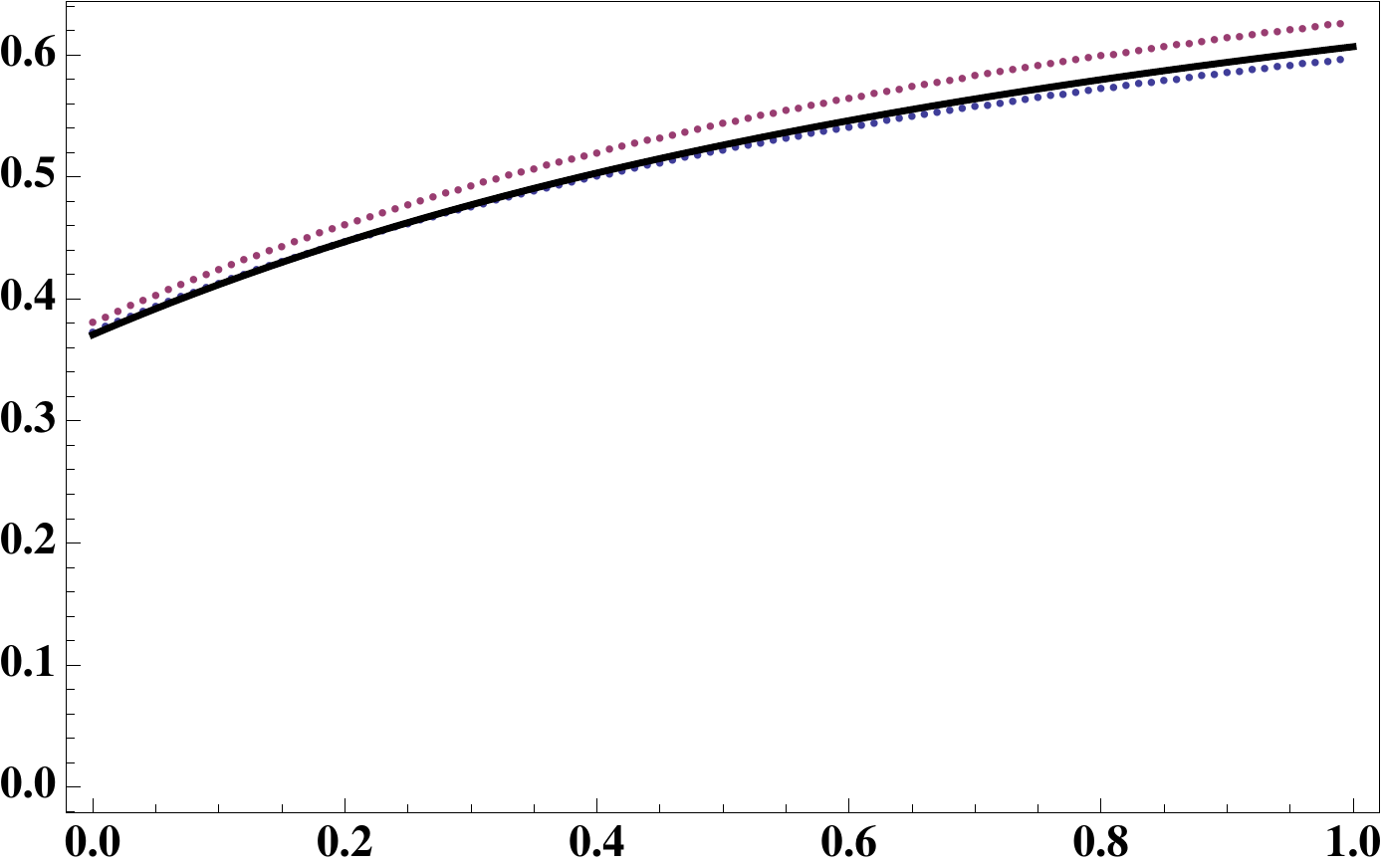}
\put(165,-10){\Large{\bf{$\alpha$}}}
\put(-25,105){\Large{\bf{$W$}}}
\put(230, 20){\includegraphics[scale=0.8]{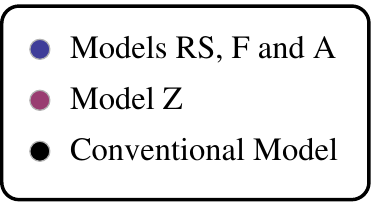}}
\end{overpic}\\\vspace{2em}
\caption{\small{Expected mean fitness $W$ as a function of the rate of self-fertilisation $\alpha$ calculated numerically and extended over 5000 loci with a mutation rate $\mu = 10^{-4}$ (giving a genomic mutation rate of $0.5$), $s = 0.1$ and $h = 0.2$. Expectations from the conventional population genetics model is calculated using equations \ref{load} and \ref{qck}.}}
\label{figL}
\end{figure}

 In the presence of self-fertilisation, selection against deleterious mutations is not as efficient as can be seen from the equations for recessive mutations in Table \ref{Lmut} (i.e. the higher the coefficient of selection $s$, the greater the differences between continuous and discrete time models). Though the introduction of self-fertilisation leads to higher frequencies $q$ of the deleterious allele than in discrete time models, there are no or very little consequences on the genetic load. This remains true even when selection is relatively strong (s = 0.1) at a large number of independent loci  (see Figure \ref{figL}).

\subsection{Genotypic frequencies and inbreeding depression}
\par
\begin{figure}
\center
\begin{overpic}[scale=0.82]{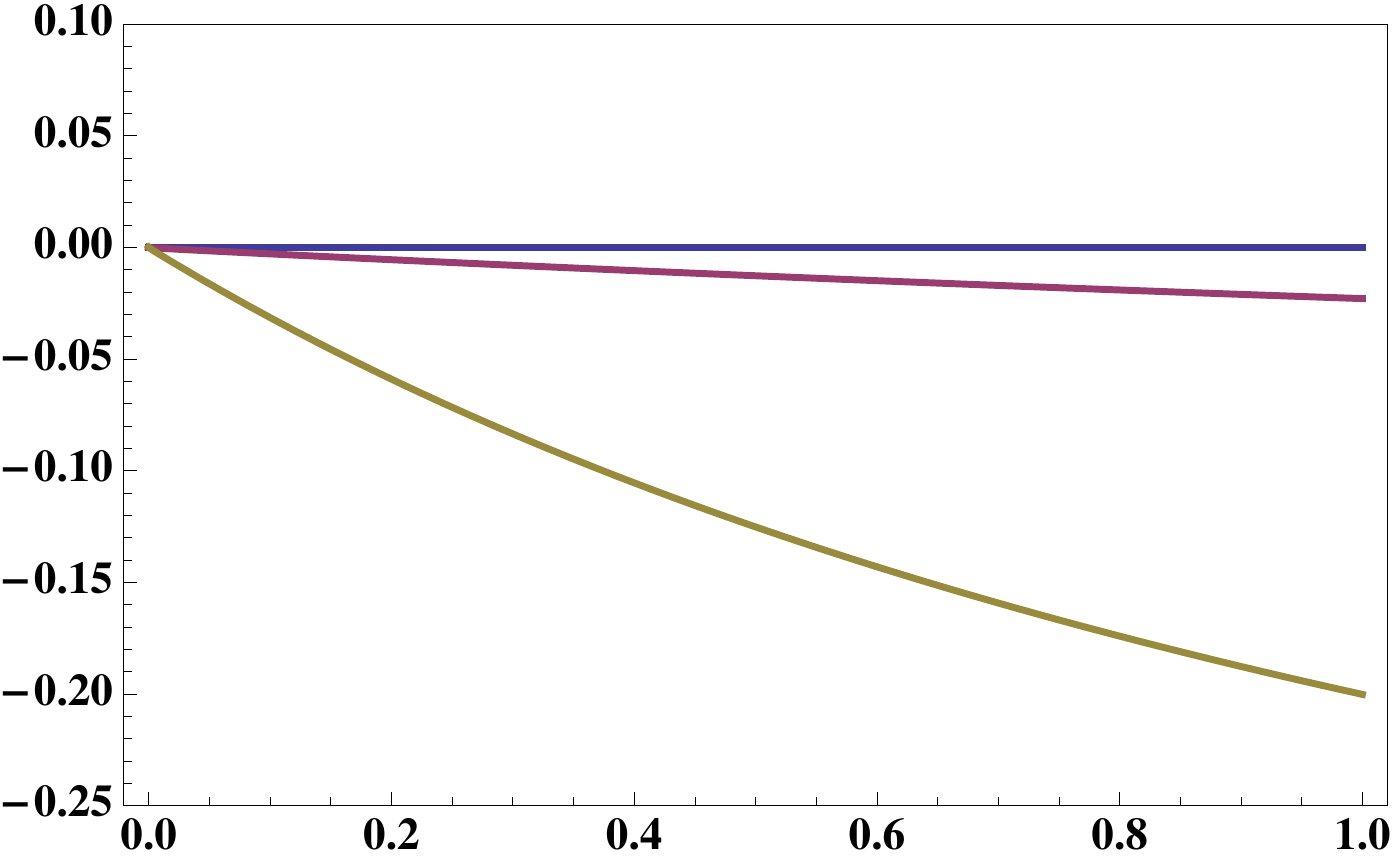}
\put(178,-10){\Large{\bf{$s$}}}
\put(-25,30){\rotatebox{90}{\bf{Excess of heterozygotes}}}
\put(40, 20){\includegraphics[scale=0.8]{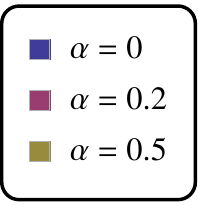}}
\end{overpic}\\\vspace{2em}
\caption{\small{Excess of heterozygotes in continuous time compared to discrete time (using equation \ref{Fexcess}) for selection models RS, F and A as a function of the coefficient of selection $s$ when mutations are recessive ($h = 0$) and $\mu = 10^{-5}$ (but these results hold for any $\mu < s$). Exact solutions obtained in the case of $h = 0$ were used.}}
\label{Fst}
\end{figure}

 In spite of genotypic frequencies for discrete and continuous time models being equivalent in the case of neutral alleles with no mutation, in the presence of both mutation and selection, this is no longer the case. As can be seen in Table \ref{Lmut}, the frequency at equilibrium $q$ of the deleterious allele in the presence of self-fertilisation when mutations are recessive ($h = 0$) is higher for the model in continuous time compared to the discrete-time model. If in continuous time, as in discrete time, the expected frequency of deleterious homozygotes in a discrete time population is $q^2(1-F_i)+ q F_i$, then if $g$ is higher, than the frequency of the deleterious homozygote $X$ should also be higher. This however is not the case, as the explicit expression for the frequency of the deleterious homozygote $X$ is the same as that obtained in discrete time. The higher frequency $q$ is therefore not associated with a greater frequency of homozygotes, but with a higher frequency of heterozygote individuals. Indeed as can be seen in Figure \ref{Fst}, there is an excess of heterozygotes in our model compared to conventional discrete-time models (shown by the negative $F_{excess}$), which is accentuated for higher rates of self-fertilisation and higher coefficients of selection. The timing of selection also influences the inbreeding coefficient, with model Z leading to even larger negative values of $F_{excess}$ than the other models (results not shown). For all continuous time models, contrary to discrete-time non-overlapping generations, Wright's inbreeding coefficient $F_W$ (equation \ref{FW}) is null, indicating that the observed frequency of heterozygotes is that expected in completely outcrossing populations.

\par
\begin{figure}
\center
\begin{overpic}[scale=0.85]{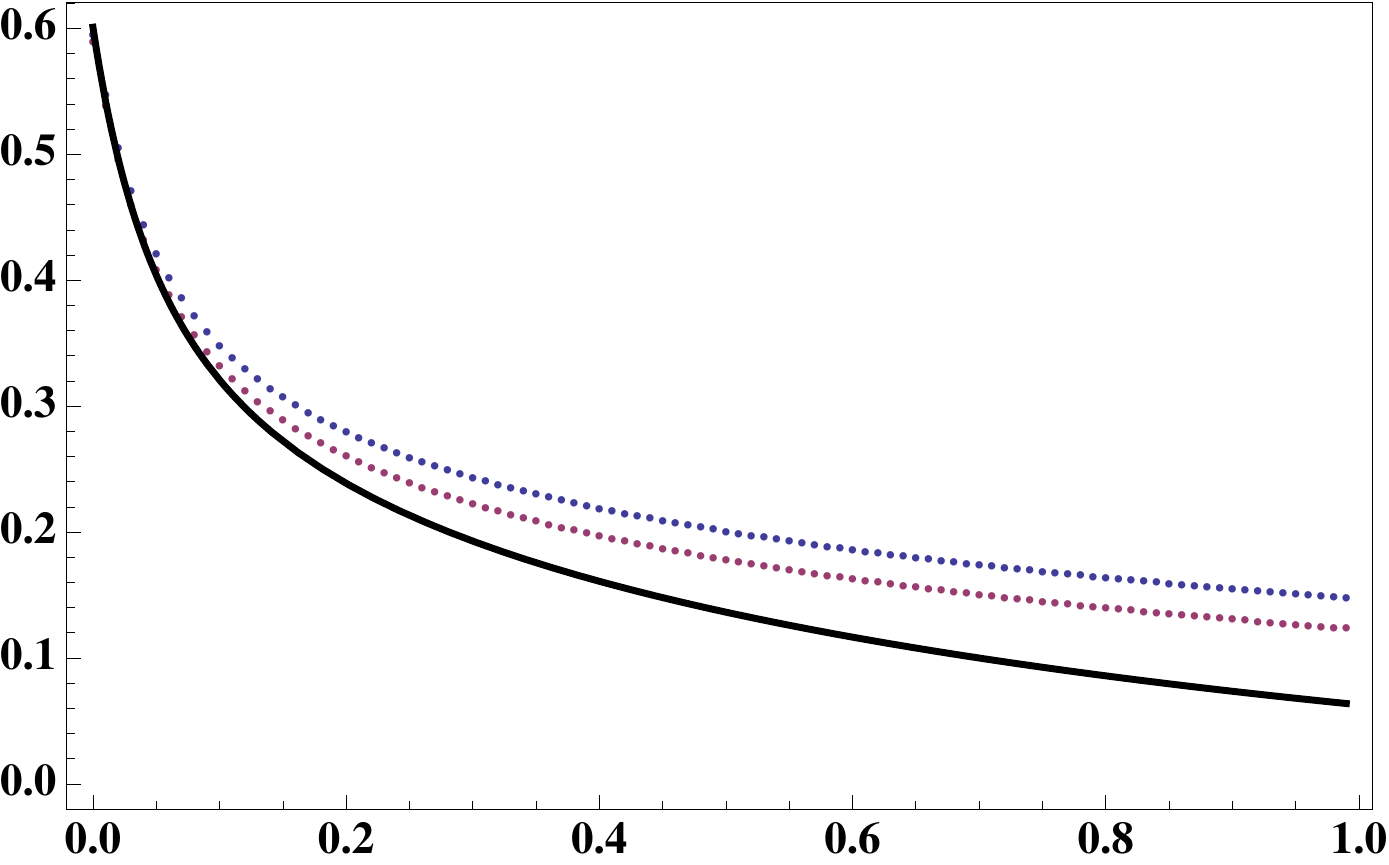}
\put(165,-10){\Large{\bf{$\alpha$}}}
\put(-25,105){\Large{\bf{$\delta$}}}
\put(240, 155){\includegraphics[scale=0.8]{Legend.eps}}
\end{overpic}\\\vspace{2em}
\caption{\small{Expected mean level of inbreeding depression $\delta$ as a function of the rate of self-fertilisation $\alpha$ calculated numerically and extended over 20 000 loci with a mutation rate $\mu = 10^{-5}$ (giving a genomic mutation rate of $0.2$), 3000 loci have $s = 1$ and $h = 0.02$ and 17000 with $s = 0.001$ and $h = 0.2$.}}
\label{figdelta}
\end{figure}

 The observed excess of heterozygotes, though it has little effect on the genetic load $L$, leads to an increased level of inbreeding depression $\delta$ within the population (see expressions for $\delta$ in Table \ref{Lmut}). As illustrated in Figure \ref{figdelta}, the level of inbreeding depression is not as easily purged in continuous time as it is in discrete time. This is all the more true in the case of very deleterious mutations. As can be seen in Figure \ref{figdelta}, model $Z$ leads to a lower observed $\delta$ due to the lower frequency of the deleterious allele.

\subsection{Population size}
From our results in the previous section, if mean fitness suffices in explaining the observed population size, then we would expect the same population size for selection on Reproductive Success, Fecundity and Adult Survival, whereas selection on Zygote survival would lead to a larger population size (as load is slightly smaller at equilibrium). This however is not the case, as can be seen in Figure \ref{figNL} and Table \ref{Nmut} (see Supplementary Material S4 for proofs of population size at equilibrium without self-fertilisation and recessive mutations). Selection on Adult Survival is the only case where population size is unaffected by the introduction of recurrent deleterious mutations ($N_{mut} = N_{eq}$ for any genetic parameters even in the presence of self-fertilisation) and so will not be mentioned further in this section. 
\par In the case of panmixia (see Table \ref{Nmut}), the expected genetic load from conventional population genetics models appears in the explicit solutions. In spite of differences in the observed mutation load for models F and Z (see Table \ref{Lmut}), they present the same population size, directly proportional to the observed mean fitness for model F. Selection on Reproductive Success results in the largest effect of mutations on population size as it is influenced by the square of mean fitness (Table \ref{Nmut}).

\begin{table}[htbp]
\center
{\renewcommand{\arraystretch}{1.75}
\begin{tabular}{|l|c|c|}
\cline{1-3}
Model&$\alpha = 0$&$h = 0$ \\ [0.2cm]\cline{1-3}
Reproductive Success&$N_{eq} (1- L^{CK}_0)^2$&$N_{eq}((1-\mu)^2 + \alpha \mu(1-\mu))$\\ [0.2cm]\cline{1-3}
Fecundity&\multirow{2}{*}{$N_{eq}(1-L^{CK}_0)$}&\multirow{2}{*}{$N_{eq}(1-\mu)$}\\ [0.2cm]\cline{1-1}
Zygote Survival&&\\ [0.2cm]\cline{1-3}
Adult Survival&$N_{eq}$&$N_{eq}$\\ [0.2cm]\cline{1-3}
\end{tabular}
}\caption{Explicit solutions for expected population size in a panmictic population ($\alpha = 0$) and all values of $\mu$, $s$ and $h \neq 0.5$ or a population with selfing rate $\alpha$ and $h = 0$. $L^{CK}_0$ is given by setting $F_i = 0$ in equation \ref{qck} which is then injected into equation \ref{load}.}
\label{Nmut}
\end{table}

\par Though the rate of self-fertilisation does not influence mean fitness in the case of recessive mutations (Table \ref{Lmut}), it can influence population size when selection is on Reproductive Success. Results from numerical iterations confirm that the relationship between mean fitness and population sizes in the case of selfing and completely recessive mutations (Table \ref{Nmut}) remain valid for mutations with any dominance, $\mu$ (the genetic load when mutations are recessive) can simply be replaced by $L$ obtained for any genetic parameters. We find that, as for mean fitness, increasing the rate of self-fertilisation increases population size for models RS, F and Z (see Figure \ref{figNL}). The rate of self-fertilisation $\alpha$ has a greater impact on population size for model RS, though for strictly self-fertilising populations ($\alpha = 1$), models RS, F and Z all result in the same population size. 

\begin{figure}
\center
\begin{overpic}[scale=0.9]{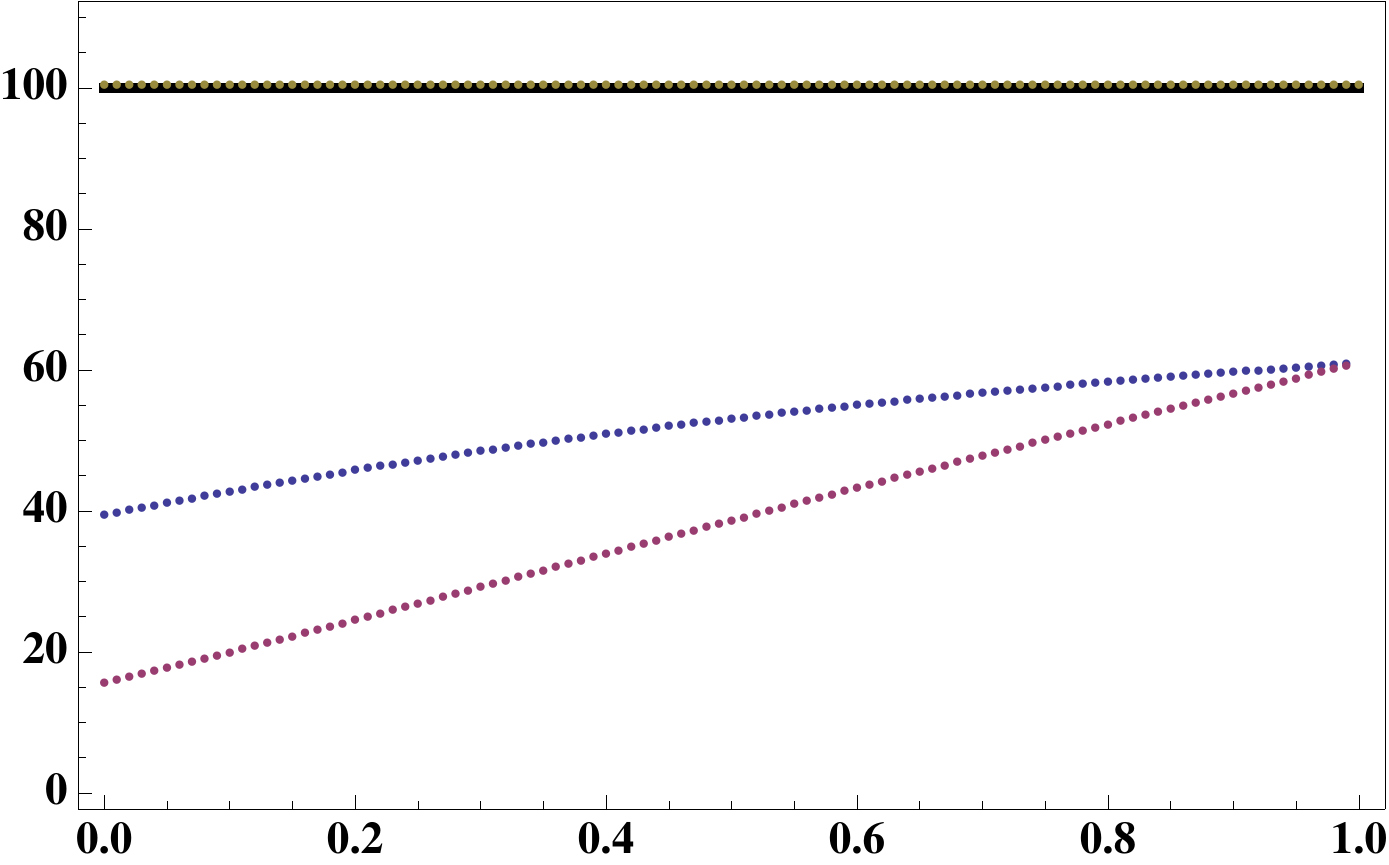}
\put(187,-10){\Large{\bf{$\alpha$}}}
\put(-25,100){\Large{\bf{$N_{mut}$}}}
\put(275, 25){\includegraphics[scale=0.8]{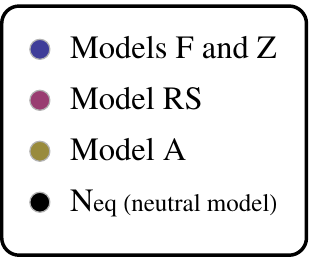}}
\end{overpic}\\\vspace{2em}
\caption{\small{Expected population size at equilibrium $N_{mut}$ as a function of the rate of self-fertilisation $\alpha$ calculated numerically and extended over 5000 loci with a mutation rate $\mu = 10^{-4}$ (giving a genomic mutation rate of $0.5$), $s = 0.01$ and $h = 0.2$. Demographic parameters are set to $b = 1$, $q = 1$ and $K = 100$, giving $N_{eq} = 100$, represented by the black line.}}
\label{figNL}
\end{figure}

\section{Discussion}
 \par Through this work, we explore whether populations in continuous time present similar genetic properties at mutation-selection balance as expected from models in discrete time in order to evaluate the potential consequences of self-fertilisation when considering a perennial life-history. We test whether different components of fitness that seem interchangeable in conventional discrete-time population genetics models with non-overlapping generations influence genetic equilibria in the presence of non-random mating. Taking this work one step further, we also take into account the potential consequences of selection against recurrent deleterious mutations on population size in a simple ecological context in order to determine how the components of fitness interact with the mating system considered.
 
\subsection{Consequences of continuous time on genetic variables}
\par Historically, continuous time models in population genetics models follow Fisher's Fundamental Theorem \cite{Fisher30}. The general consensus is that, despite a small deviation from Hardy-Weinberg proportions (accentuated by strong selection,\cite{Charlesworth70}), continuous and discrete time models are equivalent when it comes to the genetic state of populations. This is indeed the case when mating is random, but as stated in \cite{Nagylaki74}, this would not necessarily hold in non-random mating populations. As the interest of this paper is to examine equilibrium properties of populations, we did not opt for using Malthusian fitnesses as is often the case in continuous time models (since at equilibrium, the Malthusian fitness is the same for all genotypes, \cite{Charlesworth70}), but preferred a form closer to that used in discrete population genetics models, allowing for an easier comparison. 
\par We find that introducing non-random mating in a continuous time model not only leads to deviations from Hardy-Weinberg proportions, but can also change the expected frequency of the deleterious allele. In continuous time we find a higher frequency $q$, leading to an excess of heterozygotes than expected for a given rate of self-fertilisation in discrete models. 
\par  Indeed, it has been shown by Spigler et al. \cite{Spigler2009} that heterozygocity may persist in perennial populations in spite of increased inbreeding for loci under selection (heterozygocity was measured using enzyme systems) as there was an excess of heterozygotes and a lower than expected frequency of homozygotes. Furthermore, a null inbreeding coefficient $F_W$ (heterozygous frequencies in the presence of inbreeding are the same as observed in the absence of inbreeding) have also been observed in natural populations, where at the adult stage the coefficient of inbreeding was null in spite of non-null levels of self-fertilisation \cite{Duminil2016}. In these empirical works the authors concluded that this was due to stronger selection against inbred individuals. Our results indicate that, in the presence of self-fertilisation, the higher frequency of heterozygotes is not necessarily due to stronger selection. Selection on Zygote Survival (model Z) does indeed allow for a better purge of deleterious alleles (homozygotes do not arrive at the adult stage as frequently), and agrees with Duminil et al.'s \cite{Duminil2016} observations of lower survival from seedling to adult stages for inbred individuals. However, $F_W$ is null for all of our models of selection. Selection at the adult stages results in the same strength of selection against homozygotes as in discrete time (the frequency of homozygotes in continuous time is the same as that observed in discrete time for recessive mutations, regardless of the higher $q$). We therefore propose that their observations are due to a life-history with overlapping generations. A high fitness individual has a higher probability of becoming an established adult (models Z and A), and a higher probability of reproducing (models RS and F). Once established, all adults have the same mortality. Deleterious mutations are therefore continuously introduced via mutations in the offspring of unmutated (or more fit) individuals, accounting for the increase in the observed frequency of heterozygotes. 
\par Self-fertilisation is usually associated with a decrease in the frequency of heterozygotes \cite{ChD03, Glemin03} and hence a lower level of inbreeding depression (inbreeding when most individuals are homozygous has little effect on the fitness of offspring, \cite{BatKirk2000}). Both selection at the adult and zygote stages lead to higher levels of inbreeding depression $\delta$ on the population level. Since inbreeding depression is a relative measurement, the value of $\delta$ depends on the partners available in a population. When a population is made up of homozygous individuals, a population's level of inbreeding is null. Increasing the frequencies of heterozygotes, reproduction with one's self is less advantageous than outcrossing, since self-fertilising heterozygotes will automatically produce $\frac{1}{4}$ deleterious homozygotes, while outcrossing between heterozygous and homozygous wild-type individuals does not result in the production of deleterious homozygotes. The higher levels of inbreeding depression observed in our continuous time model are therefore directly related to the higher frequencies of heterozygotes observed on the population level. Selection at the zygote stage leads to a slightly lower $\delta$ due to the more efficient purge of the deleterious allele.

\subsection{Demographic consequences of the genetic load}
\par Previous models studying the effect of the genetic load on population size have done so by either decoupling the genetic and demographic models \cite{Clarke73,AgraWhit2012}, have focused on density-dependent selection with no mutation \cite{Charlesworth71, roughgarden1979}, or if selection was density independent, it occurred between the zygote and adult stage \cite{Charlesworth71}. Our main goal is to provide a general framework in which both the genetic and demographic properties of populations at mutation-selection balance are explicitly taken into account in order to evaluate whether the decoupling of genetics and demography in previous models is justified. In this work we have therefore allowed for both the genetic and demographic variables to be emerging properties. While our results concerning the genetic composition of populations hold only for continuous time, the fact that we obtain expressions for population size at equilibrium as a function of the genetic load implies that the demographic properties should be valid for both continuous and discrete time.
\par In agreement with existing models \cite{Wallace1970, Charlesworth71, Clarke73, AgraWhit2012} population size depends greatly on the timing/form of selection considered. As shown in previous models when selection occurs at a stage where resources are not wasted, then population size is not affected \cite{AgraWhit2012}, which in our case is that with selection on adult survival (model A). The assumption made is that non-viable adults are considered to be present for a sufficient amount of time to be censused, but not to compete for resources or mates with viable, reproducing adults. All other models of selection show a direct link between the genetic load and population size (as in equation 17.1b of \cite{roughgarden1979}). 
\par Selection on Fecundity and Zygote Survival (models F and Z respectively) both lead to the same equilibrium population size as in both cases, adults that produce less offspring (either because they have a lower reproductive capacity or because their offspring are not viable) take up the same amount of resources as adults with a high reproductive capacity. This creates a lag, even though we did not explicitly introduce this (as for example in \cite{AgraWhit2012} where resources are considered available at a certain rate). For model Z it is the load at the zygote stage, and not the observed load in the adult, that defines the effect on population size. This is intuitive enough as resources are invested in producing non-viable zygotes at a rate equivalent to the genetic load before selection has acted.
\par Reproductive success decreases a population's demographic output by the square of it's fitness. This is because the genotype of both parents must be taken into account when producing a zygote. It is therefore the fitness of the mating pair, which we consider to be multiplicative, that determines the outcome of a reproductive event. The extreme case would be considering a highly fertile individual mating with a sterile individual. This phenomenon (the indirect effect of a partner's fitness on one's own productivity) has been suggested by Haldane \cite{Haldane37} where on the potential loss of fitness in populations he stated ``For example, the fitness of an unfit type is generally lowered by inbreeding, because it is more likely to find an unfit mate than in an outbred population'' (though in our case, whether the population is inbred or not is not an issue).

\subsection{Implications for the evolution of self-fertilisation}
\par When comparing genetic variables at mutation-selection balance from discrete and continuous time models in the presence of self-fertilisation, we find that there are two important differences. First, the purge of deleterious mutations is less efficient at higher rates of selfing and stronger coefficients of selection, and second there is a higher frequency of heterozygotes than expected in discrete time models. Though this has little effect on the genetic load (see Figure \ref{figL}), the level of inbreeding depression does not decrease as much with self-fertilisation as it does in discrete time (see Figure \ref{figdelta}). Indeed, even though conventional population genetics models predict low levels of inbreeding depression in selfing populations, our results concord with empirical observations of high levels of inbreeding depression in highly selfing species \cite{Winn11}. When considering exclusively perennial species, the general consensus is that they maintain higher levels of inbreeding depression, though it remains unclear why this is so, though both demographic and genetic hypotheses have been proposed \cite{Morgan97}. The main implication of this result is that if overlapping generations lead to the maintenance of higher levels of inbreeding depression, self-fertilisation should not evolve as easily in perennial populations (as a transition from an outcrossing to self-fertilising regime requires that the level of inbreeding depression be below $0.5$,\cite{ChCh87}).
\par Using our definitions of fitness, an unfit individual that arrives to adulthood has as much probability of remaining a part of the population and contributing to its genetic load and inbreeding depression as a fit individual. There is no need to assume that lower reproductive capacity is necessarily accompanied by a lower viability \cite{Orr2009}. Our results therefore differ from previous theoretical works \cite{Morgan2001} where perenniality was accompanied by a more efficient purge of deleterious mutations. Deleterious mutations that do not affect longevity are more easily maintained in such a population, with deleterious mutations acting at early life stages being found at a reduced frequency among the surviving adults.
\par Though the demographic and genetic assumptions presented in the model are very simplistic, our results imply that self-fertilisation is not only genetically advantageous but can also be so demographically. The more efficient purging of deleterious mutations compared to outcrossing \cite{ChD03,Glemin03,Wright13} leads to a lower genetic load and a higher population size in general. Due to this lower genetic load, population size, if affected by load, is then expected to be higher than for outcrossing populations. However, for populations with equivalent load, selection on traits directly influencing reproductive success (not the quantity of gametes but the efficiency in producing zygotes), self-fertilisation could still lead to an increased population size if selection decreases performance in a trait linked to reproductive success. This is because a very fit individual would risk lowering its reproductive success if it reproduces with another potentially unfit individual making it less risky to self-fertilise. Selection on such traits could further facilitate the evolution of self-fertilisation by increasing an individual's reproductive success if it self-fertilises not because of producing purged offspring, nor by providing reproductive assurance \cite{Baker55}, but simply because an individual's reproductive output has not been ``diluted'' by less fit mates. 
\par Our results support that more general models combining both genetics and demography, allowing for the exploration of different modes of selection and life-histories (discrete v.s. continuous time) can provide insight into the ecological and evolutionary consequences of the genetic load.

\section*{Acknowledgements}
A special thanks also goes out to Clotilde Lepers for our inspiring discussions. This work benefited from the support of the Chair ``Modélisation Mathémathique et Biodiversité" of Veolia Environnement - École Polytechnique - Muséum National d’Histoire Naturelle - Fondation X, the French Agence Nationale de la Recherche Mod\'eles Al\'eatoires en \'Ecologie, G\'en\'etique et \'Evolution (MANEGE; ANR-09-BLAN-0215) and the French Agence Nationale de la Recherche ANR-11-BSV7- 013-03. Viet Chi Tran was supported in part by the Labex CEMPI (ANR-11-LABX-0007-01).

\section*{References}
\bibliographystyle{abbrv}
\bibliography{REFS.bib}

\end{document}